\documentclass{article}
\input psfig
\begin{document}
\title{\bf Gravoelectromagnetic approach to the gravitational Faraday 
rotation in stationary spacetimes }
\author{Mohammad Nouri-Zonoz \\
\\ 
Institute of Astronomy, Madingley Road, Cambridge CB3 0HA 
\thanks{mnzonoz@ast.cam.ac.uk}\\
\&\\
IUCAA, Post bag 4, Ganeshkhind, Pune 411 007, India
\thanks{nouri@iucaa.ernet.in}}
\maketitle

\begin{abstract}
Using the $1+3$ formulation of stationary spacetimes we show, in the 
context of gravoelectromagnetism, that the plane
of the polarization of light rays passing close to a black hole 
undergoes a
rotation. We show that this rotation has the same integral form as 
the usual Faraday effect, i.e. it
is proportional to the integral of the component of the gravomagnetic 
field along the propagation path. We apply this integral formula 
to calculate the Faraday rotation induced by the Kerr and NUT
spaces using the quasi-Maxwell form of the vacuum Einstein equations.
\end {abstract}

PACS number(s): 04.20.-q, 04.70.Bw

\section{Introduction}
It is a well known fact that the plane of the polarization of light 
rays passing through plasma in the presence of a magnetic 
field undergo a rotation
which is called Faraday rotation (Faraday effect) [10]. One can 
show that a
plane polarized wave is rotated through an angle $\Delta \theta$ given by
$$\Delta \theta = {2\pi e^3 \over m^2 c^2 \omega ^2}\int_a^b nB_{||}
dl\eqno (1)$$
where $B_{||}$ is the component of the magnetic field along the
line of sight.\\
It is also a well known consequence of the general relativity that light
rays passing a massive object are bent towards it. Several authors have
considered
 the gravitational effect on the polarization of light rays by analogy
with the Faraday effect [2,4,9]. In particular they have considered the 
propagation of electromagnetic waves
in the Kerr spacetime. In [4] the authors have used the
Walker-Penrose constant to calculate this effect for a Kerr black hole .
They have shown that in the weak field limit the rotation angle of the
 plane of the polarization is proportional to the line-of-sight component
 of the black hole's angular momentum at the third
order. In what
follows we will use the Landau-Lifshitz 1+3 splitting of stationary 
spacetimes and show that the gravitational Faraday rotation has the same
integral form as the usual Faraday effect if one replaces the magnetic 
field with the gravomagnetic field of the spacetime under consideration.
Having found
this integral form, one can use the quasi-Maxwell form of the vacuum 
Einstein equations to calculate the effect much more easily. 
In particular we  show
that the gravitational Faraday rotation in NUT space is
zero, a result which needs a lot of calculation if one uses the 
approach based on the Walker-Penrose constant.
\section {1+3 formulation of stationary spacetimes (projection formalism)}
Suppose that $\cal M$ is the
4-dimensional manifold of a stationary spacetime with metric 
$g_{ab}$\footnote {Note that the Roman indices run from $0$ to $3$ 
and Greek indices from $1$ to $3$.} and
$p\in \cal M$, then one can
show that there is a 3-dimensional manifold $\Sigma_3$
defined  invariantly by the smooth map [3]
$$\Psi :{\cal M} \rightarrow \Sigma_3$$
 where $\Psi=\Psi (p)$ denotes the orbit of the timelike
Killing vector $\mbox{\boldmath $\xi$}_t$ passing through $p$. 
The 3-space
$\Sigma_3$ is called the factor space $ {\cal M} / {G_1}$,
 where $G_1$ is the 1-dimensional group of transformations generated by
$\mbox{\boldmath $\xi$}_t$. Using a coordinate system adapted to the 
congruence $\xi_t = \partial_t$ we denote the projected 3-dimensional 
metric on $\Sigma_3$ by $\gamma_{\alpha\beta}$ $(\alpha, \beta =1,2,3)$. 
These are the coordinates 
comoving with respect to the timelike Killing vector. One can use 
$\gamma_{\alpha\beta}$ to define
differential operators on  $\Sigma_3$ in the same way that $g_{ab}$ 
defines
differential operators on  $\cal M$. For example the covariant 
derivative of a 3-vector $\bf A$ is defined as follows
$$A^\alpha_{;\beta}=\partial_\beta A^\alpha + 
\lambda ^\alpha_{\gamma\beta} A^\gamma$$
$$A_{\alpha ;\beta}=\partial_\beta A_\alpha - 
\lambda _{\alpha\beta}^\gamma A_\gamma$$
where  $\lambda ^\alpha_{\gamma\beta}$ is the 3-dimensional Christoffel
symbol cosntructed from the components of $\gamma_{\alpha\beta}$ in the
following way
$${\lambda_{\mu\nu}^\sigma}={1\over 2}\gamma^{\sigma\eta}(\partial_
{\nu}\gamma_{\eta\mu}+
\partial_{\mu}\gamma_{\eta\nu}-\partial_{\eta}\gamma_{\mu\nu})$$
It has been shown that the metric of a
stationary spacetime can be written in the following form [5]
$$ds^2=h(dx^0-A_\alpha dx^\alpha)^2-{dl}^2 \eqno (2)$$
where
$$A_{\alpha}\equiv g_\alpha={-g_{0\alpha}\over g_{00}}
\;\;\;\;\;\;\;\;\;\; , \;\;\;\;\;\;\;\;\; h \equiv g_{00} $$
and $${dl}^2=\gamma_{\alpha\beta}dx^{\alpha}dx^{\beta}=
(-g_{\alpha\beta}+
{g_{0\alpha}g_{0\beta}\over g_{00}})dx^{\alpha}dx^{\beta}$$
is the spatial distance written in terms of
the 3-dimensional metric $\gamma_{\alpha\beta}$ of $\Sigma_3$.
Using this formulation for a stationary spacetime one can write the
vacuum  Einstein equations in the following quasi-Maxwell form [6]
$${\rm div} \ {\bf B}_g = 0 \eqno (3)$$
$${\rm Curl} \ {\bf E}_g = 0 \eqno (4)$$
$${\rm div} \ {\bf E}_g = - \left(  \ {\textstyle {1 \over 2}}
(\sqrt{h} B_g)^2 + E_g^2 \right) \eqno (5a)$$
$${\rm Curl} \ (\sqrt{h}{\bf B}_g) =  2{\bf E}_g \times
(\sqrt{h} {\bf B}_g) \eqno (5b)$$
 $$P^{\alpha \beta} = E_g^{\alpha; \beta} + \left( (\sqrt{h}
 B^\alpha _g) (\sqrt{h} B^\beta _g) - (\sqrt{h} B_g)^2 
\gamma ^{\alpha \beta} \right) +
 E^\alpha _g E^\beta _g \eqno (6)$$
where the gravoelectromagnetic fields are
 $${\bf E}_g = - \nabla \ln h^{1/2} =
-{1\over 2} {{\bf \nabla}h \over h} \eqno(7)$$
 $${\bf B}_g =  \ {\rm Curl} \ {\bf A}. \eqno (8)$$
and $P^{\alpha \beta}$ is the 3-dimensional Ricci tensor constructed 
from
the metric $\gamma_{\alpha \beta}$. 
It is attractive to regard the combination $\sqrt{h} B_g$ ,appearing in 
the above equations, as the gravitational analogue of the magnetic 
intensity field $\bf H$ and denoting it with ${\bf H}_g$. In this way
one may think of the last term in equation (5b) as an energy current 
corresponding to a Poynting vector flux of gravitational field energy.
Note that all operations in these
equations are defined in the 3-dimensional space with metric
$\gamma _{\alpha \beta}$.
Using the timelike Killing vector of the spacetime
 one can define the above
gravoelectromagnetic fields in the following covariant forms
$${ E}_g^b = -{1\over 2}{{(\xi^a \xi_a)}^{;b}\over |\xi|^2}\;\;\;\;\;\;
\;\;\;\;\;\;\; |\xi|= h^{1/2}  $$
$${ B}_g^b = -{1\over 2}|\xi|\xi^a \varepsilon^{bcd}_a\left[
({\xi_d \over |\xi|^2})_{;c} - ({\xi_c \over |\xi|^2})_{;d}\right]$$
where $ \varepsilon^{bcd}_a$ is the 4-dimensional antisymmetric 
tensor and $'' ; ``$ denotes covariant differentiation.

\section {Derivation of the  gravitational Faraday rotation}
We use the analogy with the flat spacetime and take the plane of the
polarization of an electromagnetic wave to consist of two
3-vectors $\bf k$ and $\bf f$, the wave vector and the polarization 
vector respectively. The 4-vectors corresponding to these two 
3-vectors have the following relations
$$k^a k_a=0 \;\;\;\;\;\;\;\; k^a f_a=0 \;\;\;\;\;\;\;\; f^a f_a=1
\;\;\;\;\;\;\;\; a=0,1,2,3 \;\;\;\;\ \eqno(9)$$
Both of these 4-vectors are parallely transported along null 
geodesics [7] i.e.
$$\nabla_k k^a={\partial k^a \over \partial \lambda}+
\Gamma^a_{mn}k^n k^m=0$$
$$\nabla_k f^a={\partial f^a \over \partial \lambda}+
\Gamma^a_{mn}f^n k^m=0$$
where $\lambda$ is an affine parameter varying along the ray.
Employing an orthogonal decomposition based on the adapted coordinates,
the above 3-vectors defined on the 3-space $\Sigma_3$ can
be taken to be equivalent to the contravariant components of $k^a$ and 
 $f^a$ i.e. ${\bf k} \equiv \; ^{(3)}k^\alpha = \; ^{(4)}k^\alpha $
and ${\bf f} \equiv \; ^{(3)}f^\alpha = \; ^{(4)}f^\alpha $ [8].
One should note that the covariant counterparts of these 3-vectors are 
not the spatial components of the covariant 4-vectors $k_a$ and $f_a$
but 
$$^{(3)}k_\beta = \gamma_{\alpha\beta}\; ^{(3)}k^\alpha =\; ^{(4)}k_\beta + 
k_0 g_\beta $$ and
$$^{(3)}f_\beta = \gamma_{\alpha\beta}\; ^{(3)}f^\alpha =\; ^{(4)}f_\beta + 
f_0 g_\beta.$$ From equation (9) one can see that
the polarization vector is known up to a constant multiple of the wave
vector i.e.
both $f_a$ and $f^{\prime}_{a}=f_a + C k_a$ satisfy equation (9). 
This shows that
there is a kind of gauge freedom in choosing $f$ which enables one to put
$f_0=0$ without loss of generality and in which case
$^{(3)}f_\beta = \; ^{(4)}f_\beta$.\footnote{This choice corresponds to
 $C=-{f_0\over k_0}$ and makes $f$ orthogonal to the time lines.}
Applying the above decomposition the evolution of the 3-vectors 
$\bf k$ and $\bf f$ along
the ray is given by the spatial components of the parallel transport
equations i.e. 
$${\partial k^\alpha \over \partial \lambda}+\Gamma^\alpha_{mn}k^n k^m=0
\eqno (10)$$
$${\partial f^\alpha \over \partial \lambda}+\Gamma^\alpha_{mn}f^n k^m=0
\eqno (11)$$
Now we try to write these two equations in terms of the 3-dimensional
quantities defined on $\Sigma_3$. From equation (10) we have
$${\partial k^\alpha \over \partial \lambda}=-\Gamma^\alpha_{00}(k^0)^2-
2\Gamma^\alpha_{0\beta}k^0k^\beta-\Gamma^\alpha_{\beta\gamma}
k^\beta k^\gamma$$
to calculate this we need the following  components of the 
christoffel symbol [5]
$$\Gamma^\alpha_{00}={1\over 2}h;^\alpha$$
$$\Gamma^\alpha_{0\beta}={h\over 2}(g^\alpha_{;\beta} -
g^{;\alpha} _\beta) - {1\over 2}g_\beta h;^\alpha \eqno (12)$$
$$\Gamma^\alpha_{\beta\gamma}=\lambda^\alpha_{\beta\gamma}+
{h\over 2}[g_\beta (g_\gamma^{;\alpha}-g_{;\gamma}^\alpha)+g_\gamma
(g_\beta^{;\alpha}-g_{;\beta}^\alpha)]+
{1\over 2}g_\beta g_\gamma h;^\alpha$$
Substituting the above equations into equation (10) we have
$$ ^3 \nabla_{\bf k}{\bf k}=-k_0 ({\bf B}_g\times {\bf k}) + {k_0^2
\over h}{\bf E}_g \eqno (13)$$where
$$^3 \nabla_{\bf k} k^\alpha={\partial k^\alpha \over \partial \lambda}+
\lambda^\alpha_{\beta\gamma}k^\beta k^\gamma$$ is the 3-dimensional 
analogue of the parallel transport of $k$ along itself in $\Sigma_3$, 
and we have used equations (7) and (8) and the fact that
$$k^0={k_0\over h}+{\bf g}.{\bf k}\eqno(14)$$
Now in the same way we write equation (11) in terms of 3-dimensional
quantities defined on $\Sigma_3$ 
 $${\partial f^\alpha \over \partial \lambda}=-\Gamma^\alpha_{00}k^0 f^0
-\Gamma^\alpha_{0\beta}(f^0k^\beta+k^0f^\beta)
-\Gamma^\alpha_{\beta\gamma}k^\beta f^\gamma$$
Substituting from equations (12) and using the gauge in which 
$f_0=0$ we have
$$ ^3 \nabla_k{\bf f}=-{1\over 2}k_0 ({\bf B}_g\times 
{\bf f})\eqno (15)$$
To be able to interpret equations (13) and (15) we need to perform 
another calculation. Using the facts that ${\bf f}.{\bf f}=1$ and 
${\bf f}.{\bf k}=0$
one can show that \footnote{We have also used the relation
$${1\over 2}k_0({\bf B}_g\times {\bf f}).{\bf k}=
-{k_0^2\over h}{\bf E}_g.{\bf f}$$ which can be found by considering 
the fact that $$^3 \nabla_{\bf k}  ({\bf k.f})=(^3 \nabla_{\bf k}  
{\bf k }).\bf f+ {\bf k}.(^3 \nabla_{\bf k} {\bf f})=0.$$}

$$-{1\over 2}k_0 ({\bf B}_g\times {\bf k})={k_0^2\over h}
({\bf E}_g.{\bf f}){\bf f}-{1\over 2}k_0({\bf B}_g.{\bf f})({\bf f}
\times {\bf k})\eqno(16)$$
Using the above equation one can write equations (13) and (15) in
the following forms
$$ ^3 \nabla_{\bf k}{\bf k}={\bf L}\times {\bf k} + 
({\bf E}_g. {\bf k}){\bf k}\eqno (17)$$
$$ ^3 \nabla_{\bf k}{\bf f}={\bf L}\times {\bf f}\eqno (18)$$
where
$${\bf L}=-{1\over 2}k_0 [ {\bf B}_g-{1\over 2}({\bf B}_g.{\bf f})
{\bf f}+|\bf k|({\bf E}_g.{\bf n}){\bf f} ] \eqno(19)$$
where $\bf n={\bf f}\times {\hat{\bf k}}$ is a unit vector in
the polarization plane\footnote{The same relations were also found by
Fayos and Llosa [2] where they have not used the unit vector $\bf n$ 
that we have used here.}.
If we had only the second term in the RHS of equation (17) that 
would have
meant, by comparison with the 4-dimensional definition of the parallel
transport, that the 3-dimensional
vecor $\bf k$ is parallely transported along the
projection of the null geodesic in $\Sigma_3$ space.
But the appearance of the first
term shows that $\bf k$  has also  been rotated by an
angular velocity $\bf L$. The same rotation happens to the polarization 
vector $\bf f$ as can be seen from equation (18). Therefore the 
combination of these two equations leads us to the fact that
 the polarization plane has rotated by angular velocity $\bf L$ along
the projected null geodesic. What we are interested in is the angle 
of rotation around the tangent vector
$\hat{\bf k}$ along the path between the source and the observer. 
So we have
$$\Omega=\int_{sou.}^{obs.}{\bf L} . \hat{\bf k} \; {\rm d} \lambda$$
Substituting for $\bf L$ from (19) we have
$$\Omega=-{1\over 2}\int_{sou.}^{obs.}k_0 {\bf B}_g.\hat{\bf k}
\; {\rm d}\lambda\eqno (19a)$$
where we used the fact that ${\bf f.k}=0$, which follows from
equation (9) and the
gauge freedom which allows the choice $f_0=0$.
Now combining the following two equations
$$k_0=g_{0a}k^a =h(k^0 - g_\alpha k^\alpha)$$
$$k^a k_a=0 \equiv h(k^0 - g_\alpha k^\alpha)^2-
\gamma _{\alpha \beta}k^\alpha k^\beta =0$$
we have
$${{k_0}^2\over h}- \gamma _{\alpha \beta}k^\alpha k^\beta=0$$
or equivalently in terms of 
$k^\alpha ={{\rm d}x^\alpha \over {\rm d}\lambda}$
$${{k_0}^2\over h}=({{\rm d}l \over {\rm d}\lambda})^2 \eqno(19b)$$
finally upon substitution of (19b) in (19a) and putting
$\hat{\bf k}dl={\bf dl}$ we find
$$\Omega=-{1\over 2}\int_{sou.}^{obs.}\sqrt{h}\;
{\bf B}_g.{\bf dl}\eqno (20)$$
which has the same integral form as equation (1) i.e. the gravitational
Faraday rotation is proportional to the integral of
the component of the gravomagnetic field along the propagation path. But
 their main difference is the fact that the gravitational
Faraday rotation, given by (20), is a purely geometrical effect while the
usual Faraday effect, equation (1), depends on the frequency of the 
light ray. In the next two sections we will apply this formula to 
the cases of NUT and  Kerr black holes.
\section{Gravitational Faraday rotation in NUT space}
There is no gravitational Faraday rotation induced by NUT space and
the reason is as follows. Take a closed path  $\cal C$ around the NUT
 hole which consists of two paths (see figure 1).
Path $1$, a null geodesic which passes  close to the
black hole and path $2$ so far away that the effect of the gravitational
field (including the Faraday rotation) on the light rays is negligible
(another reason that on path 2 there is no gravitational Faraday rotation
is the fact that ${\bf B}_g \rightarrow  0$ as $r \rightarrow \infty$.).
Now using the  Stokes theorem, one can write equation (20)
in the following form
$$\Omega=-{1\over 2}\oint_{\cal C}(\sqrt{h}{\bf B}_g).{\bf dl}=
-{1\over 2}\int_{s}\nabla
\times (\sqrt{h} {\bf B}_g).{\bf dS}$$ and using equation (5b) 
 we have
$$-{1\over 2}\int_{1}(\sqrt{h}{\bf B}_g).{\bf dl}-{1\over 2}\int_{2}
(\sqrt{h}{\bf B}_g).{\bf dl}=-\int_{s}({\bf E}_g \times
\sqrt{h}{\bf B}_g).{\bf dS} \eqno(21)$$

The second term in the LHS of the above equation is zero by costruction.
On the other hand for the NUT space we have\footnote{ We have used
the following form of the NUT metric
$${ds^2}=f(r){(dt-2l cos \theta d\phi)}^{2}-f(r)^{-1}{dr^2}-(r^2+l^2)
(d\theta^2+sin^2\theta d\phi^2),$$where
$$f(r)=1-{2(mr+l^2)\over(r^2+l^2)}$$.}
$${\bf E}_g=-{1\over 2}\partial_r (ln f(r)) \hat{\bf r}$$ and
$${\bf B}_g= {2l f(r)^{1\over 2}\over r^2} \hat{\bf r}$$
Which together show that the RHS of equation (21) is also zero and 
therefore
$$-{1\over 2}\int_{1}(\sqrt{h}{\bf B}_g).{\bf dl}=0\eqno(22)$$
i.e. there is no Faraday effect on the light rays passing a NUT black 
hole close by. One can show that the same result can be obtained for
the NUT space using the approach based on the Walker-Penrose constant. 
In this
case one needs to take into account the  simplifying fact that all the 
geodesics in NUT space including the null ones lie on spatial cones [6]. 
\section{Gravitational Faraday rotation in Kerr metric}
Faraday effect in Kerr metric has already been studied and it has been 
shown
that despite previous claims [9], when light ray passes through
the vacuum
region outside rotating matter its polarization plane rotates [4]. In 
this section we will consider Two different cases:\\
1- When the orbit lies in the equatorial plane i.e. for $\theta=0$.\\
2- A more general orbit which intersects the equatorial plane and is 
symmetric about it.
\subsection{Orbits in the equatorial plane}
In this case using the definitions of ${\bf E}_g$ and ${\bf B}_g$ 
and equation (21)
 one can see that the gravitational analogue of Poynting vector 
defined by
${\bf E}_g \times \sqrt{h} {\bf B}_g$ has only one component along 
the $\phi$ direction and
therefore is normal to the plane of the orbit, which in turn leads 
to the fact that in this special case there is no
gravitaional faraday effect on light rays.
\subsection{A symmetric orbit about the equatorial plane}
In this case we need to find the orbit and we will see that one just 
needs to find the orbit in the zeroth order in $a/r$ and $m/r$
(i.e. straight line approximation)
which is done in appendix A.\\
Writing the Kerr metric in form (2) in Boyer-Lindquist coordinates
 one can see that
$${\bf A}=A_\phi={2amr{\rm sin}^2\theta\over 2mr-\rho^2}$$
from which we have
$$B_g^r={2amr{\rm sin}2\theta[2mr-r^2-a^2]\over \sqrt{\gamma}
{(2mr-r^2-a^2 {\rm cos}\theta}^2)}$$ and
$$B_g^{\theta}={2am{\rm sin^2\theta}( a^2{\rm cos\theta}^2-r^2)\over
 \sqrt{\gamma}{(2mr-r^2-a^2 {\rm cos^2\theta})}^2}$$
where$$\gamma=det\gamma_{\alpha\beta}\;\;\;\;\;\;\;\;\;\;\; {\rm and} 
\;\;\;\;\;\;\;\;\;\;\;\rho^2=r^2+a^2{\rm cos^2\theta} $$ Using the 
definition of the gravoelectric field given in (7) we have
$$E_g^r={\Delta m (a^2{\rm cos^2\theta}-r^2)\over 
{\rho^4 (\rho^2-2mr)}}$$ 
and
$$E_g^{\theta}= {rm a^2 {\rm sin2\theta}\over {\rho^4 (\rho^2-2mr) }}$$
where $\Delta=r^2+a^2-2mr$.
Substituting the above fields in equation (21) and putting
$\mu=\rm cos\theta$  we have

$$\Omega=-\int_{s}({\bf E}_g \times \sqrt{h}{\bf B}_g)_\phi{\bf dS^\phi}
=2am^2 \int_{-\mu_o}^{\mu_o}\int_{r_{orb.}(\mu)}^{\infty}
{ dr d\mu \over {(r^2+a^2 \mu ^2 -2mr)}^2}\eqno(23)$$
where $r_{orb.}(\mu)$ is the equation (of the
projection ) of the orbit in the $(r,\theta)$ plane.
To find the lowest order
Faraday effect we  calculate the above integral neglecting the $a^2/r^2$ 
and $m/r$ terms in which case we have
$$\Omega=2am^2 \int_{-\mu_o}^{\mu_o}\int_{r_{orb.}(\mu)}^{\infty}
{1\over r^4}{\rm d}r{\rm d}\mu=-{4 \over 3}am^2\int_0^{\mu_o}
{1\over {r_{orb.}}^3}{\rm d}\mu$$

Using the $(r,\theta)$
equation of the orbit given in appendix A one can calculate the
 above integral  which gives\\
$$\Omega=-{4 \over 3}am^2\int^0_{\mu_o=\sqrt{\eta}/r_{min}}
(1-{{r^2}_{min}\over \eta}\mu ^2)^
{3/2}d\mu =(1/4)\pi {\rm cos}\theta_o {am^2\over {r^3}_{min}}$$

This expression is of the third order $am^2 \over {r^3}_{min}$
which is of the same order as the result given in [4].
\section*{Discussion}
We have shown that using the 1+3 formulation of stationary spacetimes 
one can cast the gravitational Faraday rotation in exactly the same 
mathematical form as the usual Faraday effect i.e the gravitational 
Faraday rotation is proportional to the gravomagneic field of the 
spacetime along
the propagation path of the light ray. One should note that the origins 
of these two effects are completely different. The usual Faraday effect 
originates from the interactions between the electrons in plasma in one 
hand with the electromagnetic field of 
the light ray and the external magnetic field on the other hand
 and therefore depends on 
the frequency of the light ray. But the gravitational Faraday effect
is a purely geometrical one originating from the structure  of the 
spacetime
under consideration and it is normally attributed to the behaviour of 
the reference frames outside the spacetime of a rotating body. Using the 
quasi-Maxwell form of the vacuum Einstein 
equations we showed that there is an easy way to calculate the effect 
by transforming the line integral of ${\bf B}_g$ to a surface integral 
of the gravitational analogue of the Poynting vector.
More importantly the order of the effect can be seen without going 
through the detailed calculation (as in eqation (23) for the Kerr case)
 and in some cases like NUT space just a simple 
observation reveals that there is no effect at all.
For gravitational waves of small amplitude propagating
in a  curved background, one can develope the geometric optics
 in such a way that the wave and polarization 4-vectors satisfy the
same relations as given by equations (2) [7]. So it
can easily  be seen that all the main relations that we have found for 
the gravitational Faraday rotation of light rays are also applicable to
gravitational waves of small amplitude. 
\section*{Acknowledgements}
I  thank D. Lynden-Bell and H. Ardavan for useful discussions and 
comments. I would like to thank the referee for a careful reading 
of the paper and useful comments. This work was supported by the 
Ministry of Culture and 
Higher Education
of Iran through a research scholarship.
\section*{Appendix A}
The equation governing the projection of the orbit in the ($r,\theta$) 
plane for Kerr metric is given by [1]
$$\displaylines{\int^r{{\rm d}r \over \sqrt{r^4+(a^2-\xi^2-\eta)r^2+
2m[\eta+(\xi-a)^2]r-
a^2\eta}}=\cr \hfill{}\int^\theta{{\rm d}\theta
\over \sqrt{\eta+a^2cos^2\theta-\xi^2cot^2\theta}}\hfill{(A1)} \cr}$$
where $\xi$ and $\eta$ are the constants of the motion and we choose 
the case in which $\eta>0$, which corresponds to the null geodesics 
which intersect the equatorial plane and are symmetric about it
[1]. We perform the above
integrations for the case when $a/r \ll 1$ and $m/r \ll 1$ i.e. for
weak deflections
and indeed as we will see for a case in which there is no deflection in
the $(r,\theta)$ plane.
First we evaluate the LHS of the above equation which can be
written in the following form (after discarding the small terms)
$$\int{\rm d}r/r^2\sqrt{1-{r^2}_{min}/r^2}=
(1/r_{min}) {\rm arccos}(r_{min}/r)\eqno (A2)$$
where $r_{min}=\sqrt{\xi^2+\eta}$ is the leading term (in the expansion)
of the largest root of
$r^4+(a^2-\xi^2-\eta)r^2+2m[\eta+(\xi-a)^2]r-a^2\eta=0$ for small
deflection [4].\\
Now we evaluate the RHS of the equation (A1) in the same limit. This
integral can be written in the following form
$$RHS=-\int {{\rm d}\mu \over \sqrt{\eta+\mu^2(a^2-\xi^2-\eta)
-a^2\mu^4}}$$
Now using the fact that $a/r_{min}\ll 1$ and $r_{min}
=\sqrt{\xi^2+\eta}$ we can
approxiamte and evaluate the above integral as follows
$$RHS=-\int {{\rm d}\mu \over \sqrt{\eta-\mu ^2{r^2}_{min}}}
=-(1/r_{min}){\rm arcsin}(\mu \sqrt{{r^2}_{min}/\eta})\eqno(A3)$$
Equating the equations (A2) and (A3) we will have
$$r_{orb}={r_{min}\over \sqrt{1-({r^2}_{min}/ \eta){\rm cos}^2\theta}}
\eqno(A4)$$
which is the projection of the orbit in the $(r,\theta)$ plane 
for small deflections and in this case in fact no deflection 
because there is no term depending
on {\it m} or {\it a}. As one can see $r\rightarrow \infty$ when 
${\rm cos}\theta= \pm {\sqrt{\eta} \over r_{min}}$ where plus 
and minus signs correspond to the position angles $\theta_o$ and 
$\theta_s$ of the observer and the source respectively.
\pagebreak

\pagebreak
\begin{figure}
\centerline{\psfig{figure=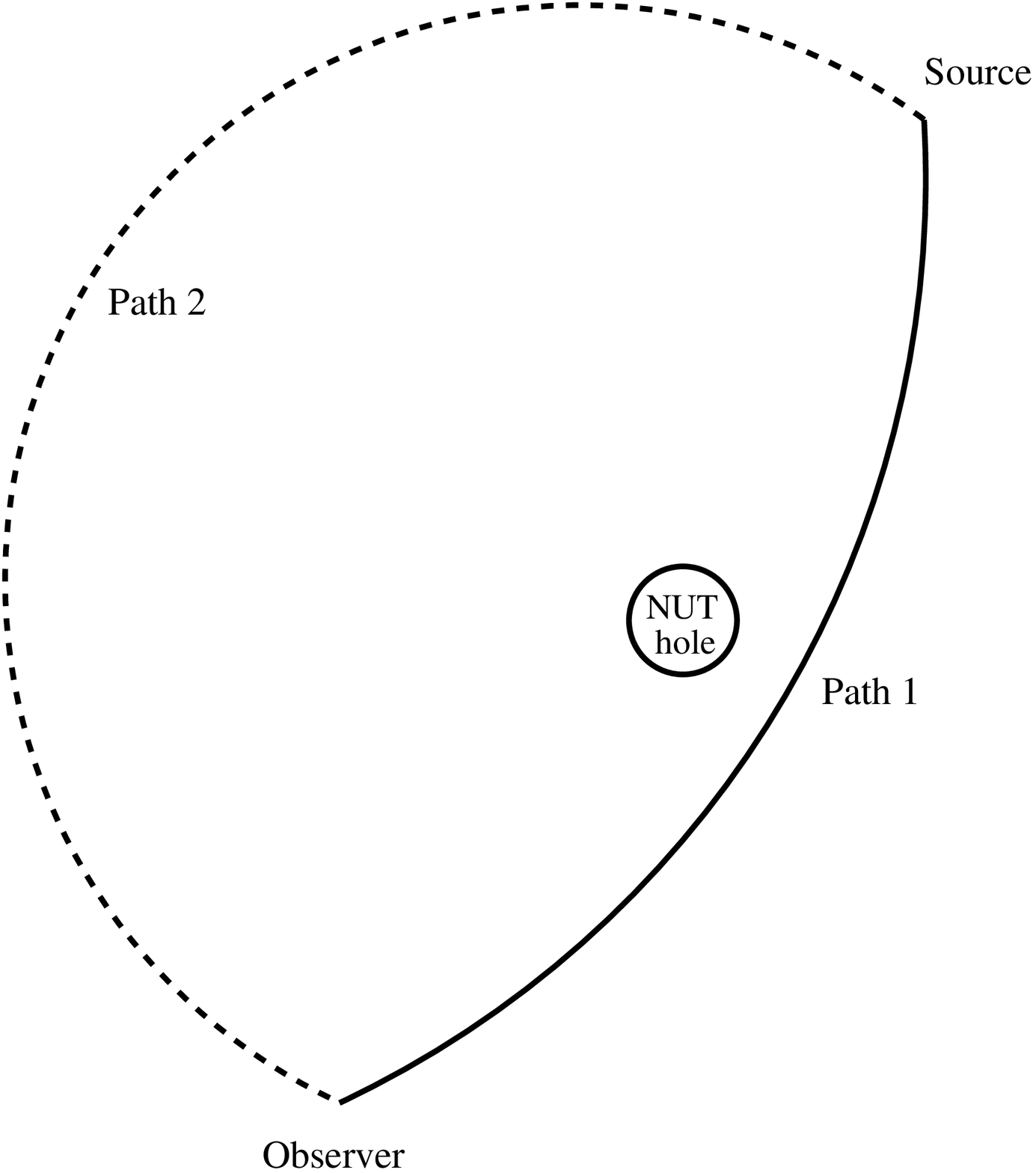,width=10cm,angle=-90}}
\caption{The NUT hole and a closed path $\cal C$ around it.}
\end{figure}

\end{document}